# Effect of magnetism on kinetics of γ − α transformation and pattern formation in iron.

**I.K. Razumov[1,2,*], Yu.N. Gornostyrev[1,2], and M.I. Katsnelson[3]**

[1]Institute of Quantum Materials Science, CJSC, Ekaterinburg, 620075, Russia

[2]Institute of Metal Physics, Russian Academy of Sciences-Ural Division, 620041 Ekaterinburg, Russia.

[3]Institute for Molecules and Materials, Radboud University Nijmegen, Heyendaalseweg 135, 6525AJ, Nijmegen, The Netherlands.

*E-mail: rik@imp.uran.ru

**Abstract.** Kinetics of polymorphous γ − α transformation in Fe is studied numerically within a model taking into account both lattice and magnetic degrees of freedom, based on first-principle calculations of the total energy for different magnetic states. It is shown that magnetoelastic phenomena, namely, a strong sensitivity of the potential relief along the Bain deformation path to the magnetic state, are crucial for the picture of the transformation. With the temperature increase, a scenario of the phase transformation evolves from a homogeneous lattice instability at $T < M_S$ ($M_S$ is the temperature of the beginning of the martensitic transformation) to the growth and nucleation of embryos of the new phase at $T > M_S$. In the latter case, a stage of formation of a tweed-like structure occurs, with a strong short-range order and slow interphase fluctuations.

## 1. Introduction

The problem of structure formation in steel related to γ-α (fcc-bcc) transition is very important for metallurgy and attracts a lot of attention [1,2]. Despite this, we are still far from a complete understanding of mechanisms and kinetics of phase transformations in steel (see Refs. [3–7]), the processes playing a crucial role in the formation of the structural state. The reason is the complexity of these nonequilibrium processes involving several spatial scale levels, from microscopic (atomistic) to macroscopic (at the level of the grain size). Concepts based on thermodynamic principles [8], can predict equilibrium properties of the system under consideration and directions of the transformations. At the same time, they are not sufficient to clarify mechanisms of the formation of a new phase, which are responsible for the morphology of this phase, depending on the regime of thermal treatment.

The mechanisms of phase and structural transformations in steel are usually categorized based on external features such as the speed of the transformation and the morphology of decay products [2]. In particular, based on the character of mass transfer the transitions are divided to diffusion and diffusionless, and, based on the way of reconstruction of the crystal lattice − to shear (cooperative) and normal, the latter occurring via nucleation and growth of noncorrelated embryos of the new phase. In practice, all transformations (except the martensitic one) involve both shear and diffusion mechanisms, their relative importance is changed with the temperature increase [9].

The current views on the nucleation of the new phase at polymorphic transformations are based on the model of thermofluctuation appearance of embryos and their consequent growth [10] what was called above the normal transformation. In this sense, the martensitic transformation [1] is very special. It represents a classical example of realization of the shear mechanism and happens at strong enough overcooling via crystal lattice instability. At the same time, the shear mechanism manifests itself also in the structure of high-temperature transformation products, like bainite and Widmanstätten ferrite [9].

Thus, understanding of physical mechanisms of the lattice instability of fcc ($\gamma$) iron is the key ingredient of our general view on the phase transformations in steel. An overwhelming majority of materials demonstrating the martensitic transformation can be treated as the Hume-Rothery alloys where a particular crystal lattice corresponds to some interval of electron concentration per atom [11]. Electronic mechanisms of the crystal lattice instabilities for these cases are reasonably well understood [12]. They are related to enhanced van Hove singularities in electron energy spectrum and to the energy gain arising when the Fermi energy lies in a pseudogap [12]. In the new crystal structure the geometry of the Brillouin zone allows to accommodate all electrons with an essential decrease of the total energy. Since the position of the Fermi energy is determined by the number of electrons per atom the Hume Rothery alloys are called also electronic phases. Typically, in these alloys the transformations are close to the second-order phase transitions with a very small hysteresis, the low-temperature phase being more closed packed than the high-temperature one. For these alloys a soft-mode picture of phonon spectra is typical [13–15]. Most of experimental and, especially, theoretical results are obtained for this kind of structural transformations. In particular, a cooperative kinetics of the transformation resulting in the formation of a coherent twin system was discussed for the cases of alloy NiAl [16] and elemental Zr [17,18].

Iron-based alloys belong to a group of rare materials where the high-temperature phase (fcc) is closed packed and the low-temperature phase (bcc, $\alpha$) is not. Neither experimental data [19] nor recent first-principle calculations [20,21] show soft-mode phonons in fcc Fe. Therefore, $\gamma \rightarrow \alpha$ transformation cannot be caused by lattice instability via the soft-mode mechanism.

As a result, an issue on the nature of pretransition state and mechanism of nucleation of the new phase at the polymorphic transformation is still a subject of hot discussions, not only for steel but also for the elemental iron and for iron-based alloys [4,15,22]. Neutron scattering data for the alloy $Fe_{70}Ni_{30}$ [23] demonstrate that atomic complexes with a strong short-range order of the martensite-phase type emerge in the premartensitic region. Therefore, one can assume that a nonclassical (shear) scheme of the phase transition is realized also for $\gamma \rightarrow \alpha$ transformation in iron-based alloys but its mechanism is more complicated than for the Hume-Rothery alloys and cannot be described in terms of individual phonon soft modes. According to Ref. [24], coupled magnetovolume fluctuations [25–30] responsible also for Invar behavior of Fe-Ni alloys can play an essential role. The situation looks paradoxical: the $\gamma \rightarrow \alpha$ transformation in iron was historically a prototype of martensitic transitions at all but this case remains still rather poorly understood, in comparison with many cases discovered later.

Starting from the seminal works by Zener [31], it is commonly accepted that magnetism plays a crucial role in the phase equilibrium of iron and its alloys, including the basic fact that bcc iron is stable at low temperatures (see, e.g, [32,33]). Magnetic and lattice degrees of freedom are especially strongly coupled in γ-Fe, which is confirmed by the results of recent first-principles calculations [22, 30, 34, 35]. As a result, magnetic ordering will be accompanied by spontaneous deformations of the crystal lattice.

As was shown in Ref. [22], the magnetic state determines not only relative energy of different phases of iron but also a character of the transformation. Fcc-Fe is stable in paramagnetic state but it looses its stability with respect to tetragonal (Bain) deformations when becoming ferromagnetic. Therefore, the martensitic scenario of $\gamma \rightarrow \alpha$ transformation emerges at the overcooling below some temperature where a strong enough short-range ferromagnetic order arises in γ-Fe. Above the critical temperature the lattice reconstruction requires an overcome of an energy barrier, and one should expect, rather, a normal transformation typical for the first-order phase transition.

A general approach to phase transitions of martensitic and martensitic-like types has been formulated in Refs. [15,36]. It is based on the expansion of Ginzburg-Landau functional for elastic energy in powers of deformations relevant for the transformation:

$$\Phi = A(e_2^2 + e_3^2) + Be_3(e_3^2 - 3e_2^2) + C(e_2^2 + e_3^2)^2 \qquad (1)$$

where the expansion coefficients are related with second-, third- and fourth-order elastic moduli, $e_1 = (\varepsilon_{11} + \varepsilon_{22} + \varepsilon_{33})/\sqrt{3}$, $e_2 = (\varepsilon_{11} - \varepsilon_{22})/\sqrt{2}$, $e_3 = (\varepsilon_{11} + \varepsilon_{22} - 2\varepsilon_{33})/\sqrt{6}$ are dilatation, deviatoric, and tetragonal deformations, respectively, $\varepsilon_{ij} = (u_{i,j} + u_{j,i} + u_{k,i}u_{k,j})/2$, $u_{i,j} = \partial u_i/\partial x_j$ is the distortion tensor, $u_i$ are atomic displacements. The third-order term in Eq.(1) provides the barrier making the transformation the first-order phase transition. Within this model, there is an evolution of mechanisms of the transformation, from the normal type (nucleation and growth) to the shear type (lattice instability) at the decrease of the parameter $B$.

The model [15,36] describes correctly main features of the shear transformation, including the formation of coherent systems of twin-like domains. However, only the components of the deformation tensor playing the role of the order parameter were taken into account in these works. As was shown in Ref. [37], *all* components of the deformation tensor should be taken into account for a proper description of elastic energy at the polymorphic transformation. As a result, intrinsic stresses turn out to be a factor affected crucially on the character of transformation. In particular, it leads to the emergence of ordered patterns, both at the developed stage of the transformation and at the pre-transition stage. It was demonstrated in Refs. [5,7,38] that the condition of compatibility of deformations (Saint-Venant's principle) results in an appearance of effective long-range interactions for the field of the order parameter. Due to these long-range effects, the transformation occurs consistently in different microvolumes and is accompanied by the pattern formation which is really characteristic of the martensitic transformation. The role of long-range interactions in the pattern formation is well known [39–41] and was discussed many times for very different systems, from stripes in high-temperature superconductors [42–44] to stripe domains in ferromagnetic films [45–47]. Within the framework of the model [5,7,38] taken into account the long-range elastic strains, it is possible to describe main peculiarities of the pattern formation at the martensitic-type structural phase transitions.

These model approaches provide us a solid background for a general understanding of polymorphic transformations. However, it is not clear how to apply them to iron and steel where there are no evidences of the soft-mode behavior of the high-temperature phase. In particular, a microscopic justification of the phenomenological expression for the free energy is absent for this case. It is worthwhile to note, also, that, according to the experimental data [48] a small enough overcooling of iron results in a shear transformation but not of the martensitic type (so called massive transformation).

Here we present a consequent model of the polymorphic transformations in iron based on the first-principles parametrization of the free energy, both lattice and magnetic degrees of freedom being taken into account. We demonstrate that the fcc lattice instability in overcooled austenite is triggered by the formation of short-range ferromagnetic order. At smaller overcooling the shear mechanism also takes place but the transformation arises via the fluctuation nucleation of the new phase and its further growth.

## 2. Formulation of the model

Here we develop the approach [5,15,36–38] and apply it to the shear transformations in iron. As in the previous works we assume the Bain mechanism of the transformation and choose the tetragonal deformation as the order parameter. We construct the expression for the free energy taken into account both lattice and magnetic degrees of freedom and perform its parametrization based on first-principles density-functional calculations.

### 2.1. Free energy functional.

We start with the Ginzburg-Landay functional which can be represented as

$$G = \int \left( g_e + \frac{k_t}{2} (\nabla e_t)^2 \right) dr, \tag{2}$$

where $g_e$ is the density of energy of lattice deformations and $k_t$ is the parameter determining the width of the interphase boundary.

We restrict ourselves to the two-dimensional case only and introduce the local density of energy similar to Refs. [5,15]:

$$g_e = g_t(e_t, T) + \frac{A_v}{2} e_v^2 + \frac{A_s}{2} e_s^2, \tag{3}$$

where $e_v = \left( \varepsilon_{xx} + \varepsilon_{yy} \right) / \sqrt{2}$ is the dilatation, $e_t = \left( \varepsilon_{xx} - \varepsilon_{yy} \right) / \sqrt{2}$ is the tetragonal deformation, and $e_s = \varepsilon_{xy}$ is the shear (trigonal) deformation, $g_t(e_t, T)$ is the energy density for the tetragonal deformation. We count the tetragonal deformation from fcc phase, so that $e_t = 0$ in γ-phase and $e_t = 1 - 1/\sqrt{2}$ in α-phase, i.e. $e_t = (1 - c/a)$, where $a, c$ are lattice periods along $x$ and $y$ directions. The coefficients $A_v, A_s$ are expressed in terms of elastic moduli [37], $A_v = C_{11} + C_{12}$, $A_s = 4C_{44}$, and $\left. \partial^2 g_t / \partial e_t^2 \right|_{e_t=0} = 2(C_{11} - C_{12})$.

The first-principles computational results for the Bain deformation path [22] allows us to find an explicit expression for the density of free energy for Fe taken into account both deformations and magnetic degrees of freedom. To this aim, we represent the magnetic-dependent part of the total energy in Heisenberg-like form

$$E = E_{PM}(\hat{\varepsilon}) - \sum_{i<j} J_{i,j}(\hat{\varepsilon}) Q_{ij}(T) \qquad (4)$$

where $Q_{ij}(T) \equiv <\mathbf{m}_i \cdot \mathbf{m}_j>$ is the correlation function of magnetic moments on sites $i$ and $j$, $E_{PM}$ is the energy of paramagnetic state, and brackets $<\ldots>$ mean average over an ensemble of magnetic configurations at a given temperature. It follows from the calculations of exchange interactions in γ-Fe that the nearest-neighbor contributions are dominant [34] and give the total exchange energy equal to $J \approx z_1 J_1$ in bcc and fcc phases ($z_1$ is the corresponding nearest-neighbor number), long-range exchange interactions are strongly oscillating and their contributions are mutually canceled. We will use therefore the nearest-neighbor approximation, so that

$$g(\hat{\varepsilon}, T) = g_{PM}(\hat{\varepsilon}) - \tilde{J}(\hat{\varepsilon}) \tilde{Q}(T) \qquad (5)$$

where $\tilde{J} = m^2 J / \Omega$ , $\Omega$ is the volume per atom and $m$ is the magnetic moment. The dependence $\tilde{Q}(T) = Q(T)/m^2$ reflects the temperature evolution of magnetic short-range order; $\tilde{Q} = 0$ for totally disordered paramagnetic state and $\tilde{Q} = 1$ for the ferromagnetic ground state. Close to the Curie temperature $T_C$ this function is strongly temperature-dependent. The exchange energy $J(\hat{\varepsilon})$ can be extracted from the computational data for the Bain path in ferromagnetic and paramagnetic states [22],

$$\tilde{J}(\hat{\varepsilon}) = g_{PM}(\hat{\varepsilon}) - g_{FM}(\hat{\varepsilon}) . \qquad (6)$$

We will assume further that the exchange energy depends on the Bain tetragonal deformation $e_t$ only, and the value of dilatation is chosen from the minimum of energy at a given $e_t$.

We approximate the temperature dependence of the spin correlator by the expression

$$\tilde{Q}(T) \equiv \frac{<\mathbf{m}_0 \cdot \mathbf{m}_1>}{m^2} = \frac{1 + \exp(-kT_C / \lambda)}{1 + \exp(k(T - T_C)/\lambda)} \qquad (7)$$

where the Curie temperature is determined by the exchange energy for a given tetragonal deformation, $kT_C(e_t) = \alpha \tilde{J}(e_t)\Omega$ with some unknown yet numerical factor α. Using the Hellmann-Feynman theorem and Eqs. (5),(7) one can represent the free-energy density with the magnetic contribution taken into account as

$$f_t(e_t, T) = g_{PM} - Ts_0 - \int_0^{\tilde{J}} \tilde{Q}(\tilde{J}', T)d\tilde{J}' = g_{PM} - Ts_0 - \frac{\tilde{J}}{\beta} + \frac{\lambda(1-\beta)}{\alpha\Omega\beta} \ln\left(\frac{\beta + \exp(\alpha\Omega\tilde{J}/\lambda)}{1+\beta}\right) \qquad (8)$$

where $\beta = \exp(kT/\lambda)$, $s_0$ includes phonon entropy and the high-temperature limit of magnetic entropy (at $<\mathbf{m}_0 \cdot \mathbf{m}_1> = 0$). The latter can be dependent on the length of magnetic moment $m$. Taking into account that both the length of magnetic moments $m$ and the phonon entropy vary quite weekly along the Bain path [22,33] we will neglect this contribution to the energetic of the transformation.

The parameter λ characterizing the width of the regions of strong short-range magnetic order was chosen within the range 0.04-0.06 eV, and the numerical factor α = 0.4 – 0.5. This gives $T_C \approx 1000$K and leads to the difference between the temperature of phase equilibrium $T_0$ and the temperature of the beginning of martensitic transformation $M_S$ in agreement with experiment, $T_0$ - $M_S \approx 250$K.

Fig. 1 shows the dependences of the total energy and free energy defined by Eqs. (5)-(8) on the tetragonal deformation for different temperatures, both above and below the Curie temperature. Curves 1 and 5 in Fig. 1a correspond to the results of first-principle calculations of the total energy with volume optimization for a given $c/a$ [22] for ferromagnetic and paramagnetic states, respectively. One can see that the shape of the curves changes essentially in a narrow temperature range at $T \sim T_C$. This demonstrates a decisive role of magnetism in energetics of $\gamma - \alpha$ transformation in Fe. It is the formation of the short-range magnetic order leads to the lattice instability and development of the martensitic transformation at temperatures far below $T_0$, that is, in overcooled austenite.

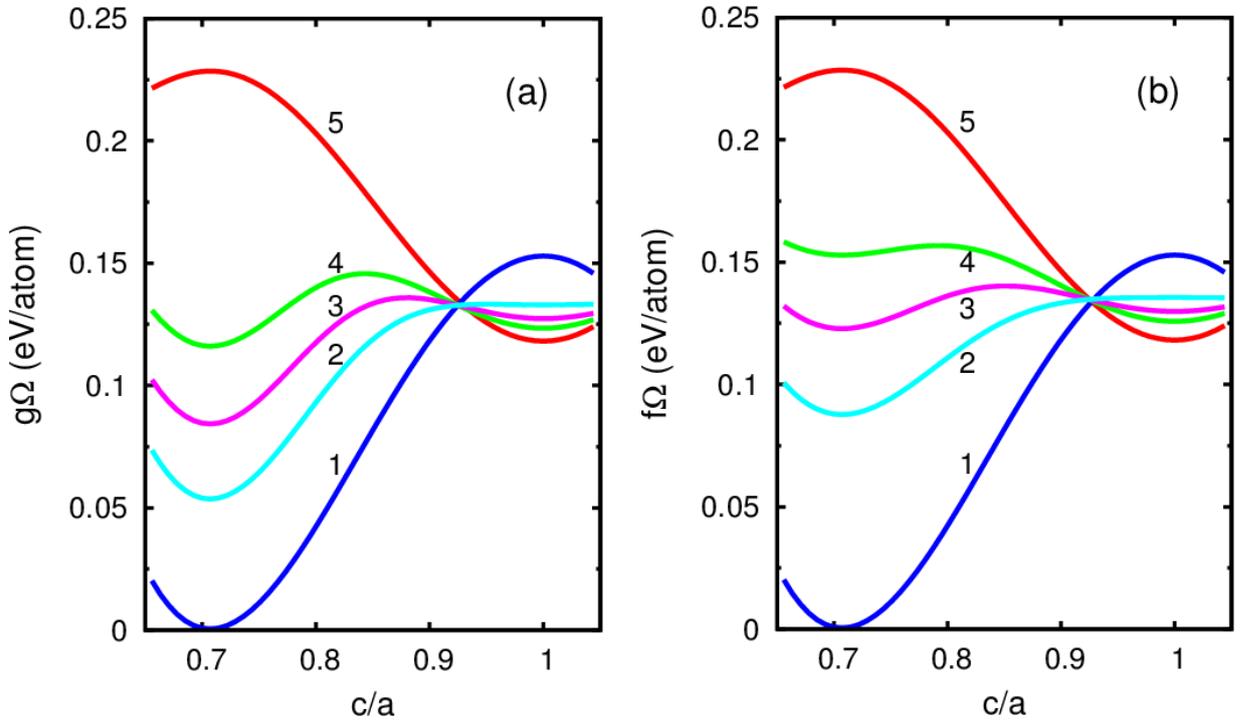

Fig. 1 (Color online). Energy (a) and free energy (b) as functions of tetragonal deformation for temperatures $T$=700K, 1000K, 1300K (curves 2, 3, 4, respectively) found from Eqs. (5)-(8) and the first-principle computational results [22] for the Bain path in ferro- (1) and paramagnetic (5) states.

The quantities $g_t$ and $f_t$ as functions on $c/a$ have two minima for temperatures close enough to $T_C$ (Fig. 1), at $c/a = 1/\sqrt{2}$ (bcc-phase) and $c/a = 1$ (fcc-phase). The magnetic entropy contribution (difference between magnetic energy and free energy) makes the energies of bcc and fcc phases closer and the barrier between them lower. The free energies of bcc and fcc phases are equal at the temperature $T_0$ ~1060K. When temperature decreases the barrier disappears at $T<M_S = 780$K. In this case the transformation will proceed as a result of spontaneous lattice instability, that is, via the martensitic mechanism.

Let us introduce the order parameter $-1 < \phi < 1$ related with the Bain tetragonal deformation as $\phi = \sqrt{2}/(\sqrt{2} - 1)e_t$. Positive and negative values of $\phi$ correspond to two possible (orthogonal) directions of the Bain deformation in two-dimensional case. It turns out that the free energy (per atom) as a function of the order parameter can be quite accurately approximated by the polynom

$$\Omega f_t(\phi) = a + b\phi^2 - \frac{(b+c)}{2}\phi^4 + \frac{c}{3}\phi^6,$$  (9)

the parameters being presented in Table 1, in the units of $\Omega\tilde{J}^\alpha$. The latter quantity, $\Omega\tilde{J}^\alpha$ =0.217eV/at characterizes the exchange interactions in $\alpha$-phase and determines a natural energy scale in the problem. One can see that in the relevant temperature interval the parameters $a$ and $c$ vary quite weakly whereas $b$ changes the sign at $T = M_S$. The values of the parameters $A_v, A_s$ where choosen as in Ref. [49]: $A_v = 97$, $A_s = 108$ (in the units of $\Omega\tilde{J}^\alpha$), whereas the parameter $k_t$ was chosen as in Ref. [37], $k_t$ =8.8·10$^{-4}$ (in the units of $L^2\Omega\tilde{J}^\alpha$ where $L$ is the sample size).

**Table 1.** Parameters $a$, $b$, $c$ (in units of $\Omega\tilde{J}^\alpha$) in the expression (9) for the free energy found from the fitting of the curves in Fig. 1.

| $T,K$ | $a$ | $b$ | $c$ |
|-------|-----|-----|-----|
| FM | 0.714 | -1.461 | -0.166 |
| 400 | 0.664 | -0.659 | 0.631 |
| 600 | 0.641 | -0.290 | 0.908 |
| 780 | 0.613 | 0.0 | 1.074 |
| 950 | 0.608 | 0.244 | 1.157 |
| 1000 | 0.604 | 0.309 | 1.171 |
| 1060 | 0.599 | 0.378 | 1.180 |
| 1170 | 0.590 | 0.502 | 1.184 |
| 1300 | 0.585 | 0.627 | 1.166 |
| PM | 0.544 | 1.226 | 0.618 |

### 2.2. Kinetic equations.

In the case of nonconservative order parameter the evolution of the system is determined by dynamic equations for atomic displacements [50]

$$\rho\frac{\partial^2 u_i(\mathbf{r},t)}{\partial t^2} = \sum_j \frac{\partial \sigma_{ij}(\mathbf{r},t)}{\partial r_j},$$  (10)

where $i,j=\{x,y\}$; $\rho$ is the mass density. A similar equation was used earlier [7] to describe the pattern formation near martensitic transformation in a model system. The components of the generalized stress tensor $\sigma_{ij}(\mathbf{r},t) = \frac{\delta G}{\delta \varepsilon_{ij}(\mathbf{r},t)}$ are:

$$\sigma_{xx} = \frac{1}{(\sqrt{2}-1)}\frac{df(\phi,T)}{d\phi} + \tilde{A}_v e_v + \tilde{A}_{vt}\left(\frac{\sqrt{2}-1}{2\sqrt{2}}\phi^2 + e_v\phi\right) - \tilde{k}_t\nabla^2\phi$$

$$\sigma_{yy} = -\frac{1}{(\sqrt{2}-1)}\frac{df(\phi,T)}{d\phi} + \tilde{A}_v e_v + \tilde{A}_{vt}\left(\frac{\sqrt{2}-1}{2\sqrt{2}}\phi^2 - e_v\phi\right) + \tilde{k}_t\nabla^2\phi$$  (11)

$$\sigma_{xy} = A_s e_s$$

where $\widetilde{A}_v = A_v / \sqrt{2}$, $\widetilde{A}_{vt} = A_{vt}(\sqrt{2}-1)$, $\widetilde{k}_t = k_t(\sqrt{2}-1)/2$.

Formulas (10),(11) give us a closed set of equations describing an evolution of the structural state during the process of polymorphic transformation via the Bain mechanism. The system (10),(11) was solved numerically at a square region with periodic boundary conditions. We used dimensionless units

$$r_i \rightarrow r_i / L,\ u \rightarrow u/L,\ t \rightarrow t \sqrt{\frac{\tilde{j}^\alpha}{L^2 \rho}},\ \rho \rightarrow 1$$ (we used in the calculations the value $L=300\Delta$, where $\Delta$ is

the unit of the spatial grid; for the chosen simulated sample size $L = 500$ nm the time unit corresponds to $2 \cdot 10^{-9}$ s). The temperature was introduced within the microcanonical ensemble. Initially, the system were heated up to high temperature by small random forces $\xi(\mathbf{r},t)$ with Poisson distribution [51]; it follows from the calculations that this way to heat the system provides the Gibbs distribution. After exposure at high temperature without random forces to reaching the equilibrium state the temperature dropped sharply to the given temperature $T$ in the 400 K to 1000 K by switching off the random forces and a rescaling of the velocity field. Further, when we will discuss temperature dependences we will mean the dependences on this final temperature, after annealing and quenching. The temperature was controlled via the average kinetic energy per degree of freedom, $kT = \rho \Omega < v^2 > /2$, where $<v^2>$ is the average (over sample) square velocity. To prevent $\gamma - \alpha$ transformation during heating and cooling the free-energy density $f_t(\phi)$ was chosen in form (9) with parameters correspond to high temperature (where only one fcc phase is stable) and switch to $f_t(\phi, T)$ for given T after achieving of the equilibrium state. To avoid a heating of the sample during transformation we cool the system via rescaling of the velocity field.

## 3. Computational results

To simulate the process of quenching of the high-temperature fcc phase we first kept the system at the temperature $T = 1200$ K, which is, for our values of parameters, above both the temperature of thermodynamic fcc/bcc equilibrium $T_0 = 1060$ K and the Curie temperature for bcc Fe $T_C = 1045$ K. The former value is a bit lower than the experimentally measured temperature of equilibrium $T_0 = 1184$ K. Taking into account simplifications made at the formulation of our model and keeping in mind that we are interested in qualitative description of the pattern formation the agreement between theory and experiment for $T_0$ seems reasonable.

Typical snapshots of distributions of the order parameter $\phi$ depending on time are shown in Fig. 2. At a strong overcooling we observe the standard picture of the martensitic transformation where a coherent system of twins of bcc phase arises in a relatively short time (Fig. 2a,c). Such kinetics takes place for the temperatures up to $M_S = 780$ K, which should be identified with the temperature of the beginning of martensitic transformation in our model. For $T < M_S$ two scenarios of the shear instability were observed. A deep overcooling ($T = 400$K, Fig. 2a) results in a homogeneous transformation which starts with a formation of two systems of mutually perpendicular standing waves. At higher temperatures ($T = 700$K, Fig. 2c) a development of fluctuations inherited at the cooling from the high-temperature state takes place. In this case, the instability proceeds in an auto-catalytic way, by a replication of a single embryo in the neighboring layers. A similar mechanism was observed earlier [38] at simulations of the martensitic transformation in a model system with an artificially introduced embryo.

To clarify the role of inhomogeneities of the order parameter inherited at a fast cooling we have carried out simulations of the transformation without a preliminary heating and quenching of the system. In this case (Fig. 2b,d) the transformation occurs from initial state containing some small fluctuations in a homogeneous way in the whole sample via emergence of s system of standing waves, with their amplitude increasing in time. At some intermediate stages, colonies of twins of $\alpha$-phase are formed, and later they are reconstructed to relax internal strains. Thus, the existence of $\alpha$-phase precursors in the state quenched from high temperatures leads to an acceleration of the transformation and appearance of rougher martensitic plates.

Within the temperature range between $M_S$ and $T_0$ the formation of the coherent twin system is preceded by a more or less long stage (the higher the temperature the longer the stage) of the tweed-like structure [37,52] (Fig. 2e) with a pronounced short range order, when domains of distorted bcc phase coexist with a distorted fcc regions. The remnants of the high-temperature fcc phase survive at this regime till the largest times used in our simulations. The character of the transformation does not change qualitatively at the temperature increase above $M_S$. This is not surprising since the energy barrier for the uniform Bain deformation (Fig. 1) is low in comparison with the temperature and the development of the transformation is decelerated by internal strains due to the nucleation of the embryo of $\alpha$-phase. In the same time, the mechanism of replication of a single embryo is less pronounced at $T > M_S$ (Fig. 2e).

At temperatures above 950 K (in our model) the transformation is not finished and the tweed structure remains the final stage (Fig. 2f). However, if one introduces embryos of bcc phase (the time instant 0.05 in Fig. 2g) the complete transition takes place leading to the formation of a metastable two-phase state (Fig. 2g). We believe that this picture corresponds to a heterogeneous nucleation of the new phase at temperatures close to $T_0$. Above $T_0$ the transition does not happen at all, although, up to the temperatures ~1200K the short-range order is clearly pronounced, as one can see in Fig. 2f.

The transformation dynamics is illustrated in Fig. 3 which shows the time-dependence of the fraction $\eta(t)$ of $\alpha$-phase regions defined by the condition $|\phi| > 0.85$. One can see that the incubation period of the transformation increases with the temperature growth (in our simulations, the maximum value $\eta < 1$, due to a contribution of interphase boundaries). Within the temperature range from $M_S$ to $T_0$ the bulk fraction of the new phase fluctuates in time and saturates due to exhaustion of the $\gamma$-phase regions, the higher the temperature the slower is the process. In the two-phase state resulting from the heterogeneous nucleation at temperatures close to $T_0$ (the curve 4 in Fig. 3), the fraction of the $\alpha$-phase demonstrates a slow dynamics (with a characteristic oscillation period of the order of $10^{-9}$ s) which exists till the largest times reached in our simulation.

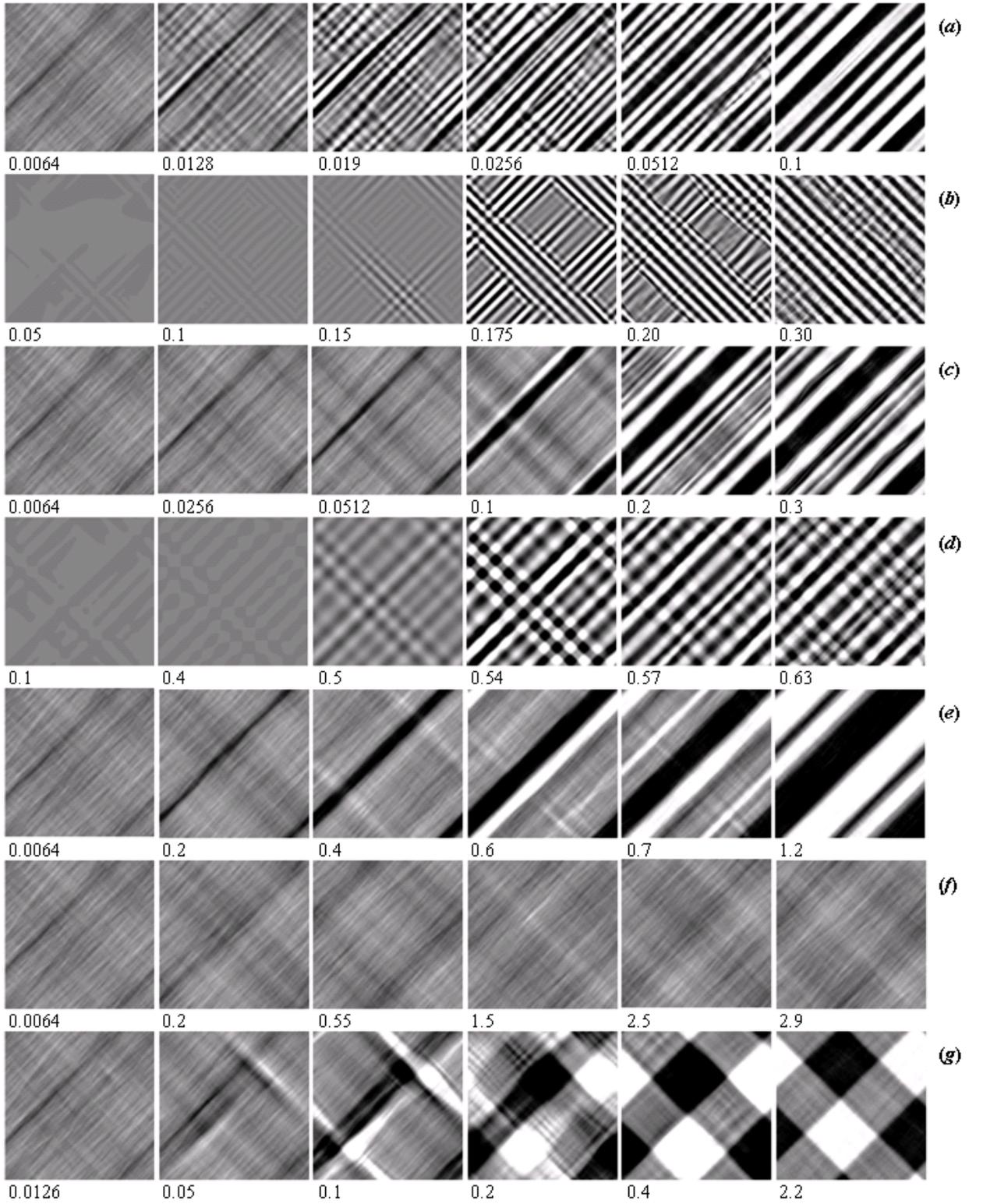

Fig. 2. Time evolution of the structure at the exposure at $T$ =400K (a,b), 700K (c,d), 950K (e), 1000K (f,g) after quenching of the high-temperature state (a,c,e,f) or development of instability of the uniform fcc state (b,d), under homogeneous (a-f) and heterogeneous nucleation (g). Gradations of grey color correspond to the value of the order parameter $\phi$; black and white colors show the regions of α-phase with two possible orientations ($\phi = \pm 1$).

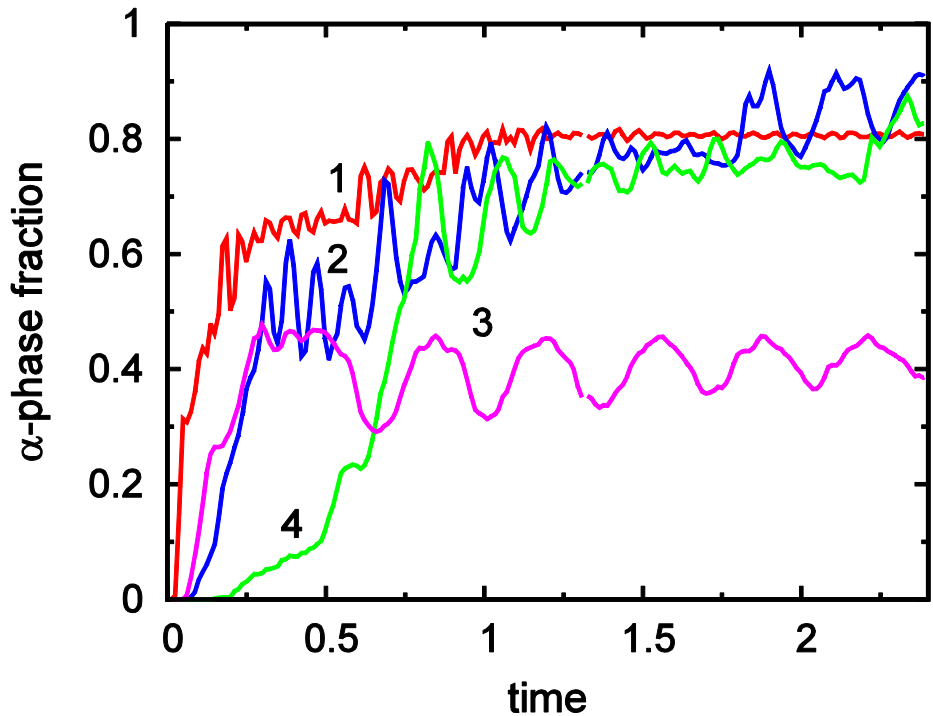

Fig. 3 (Color online). Time dependence of the fraction of α-phase under the homogeneous nucleation at $T$=400K (the curve 1),700K (2), 950K (3), and under the heterogeneous nucleation at 1000K (4).

## 4. Discussion and conclusions

Up to now, the nature of the massive transformation for the temperatures between $M_S$ and $T_0$ remains mysterious. The results presented here demonstrate two peculiarities of the massive transformation, namely, emergence of the tweed structure with a typical spatial scale of the order of tens of nanometers and a slow dynamics of fluctuations between fcc and bcc phases. Apart from this, precursors or other sources of heterogeneous nucleation play a crucial role. The last statement is not surprising, of course, but the previous two are very nontrivial. They are the main conclusions of the present work.

It is quite clear why these features were not noticed experimentally up to now. To probe the structural state at the scale of tens of nanometers one can use diffuse X-ray or electron scattering whereas the slow dynamics should be probed by the quasielastic neutron scattering. All these techniques are not very easy at high temperatures but possible.

β+ω state in titanium and zirconium alloys, e..g, Zr-Nb is the prototype example of slow (in comparison with phonon time scale) interphase fluctuations [53,54]. Their qualitative description within the framework of the simplest model was suggested in Ref. [55]. It was shown in that work that the slow dynamics arises naturally at fluctuations in a double-well potential in the case where the height of the barrier between the wells is comparable with the temperature. It is exactly the situation for the temperatures close $T_0$ in our simulations. It is important that the barrier should be flat enough; otherwise, the smallness of the barrier does not follow from the equality of energies in two minima of the potential.

In our model for iron, this flatness of the potential relief is caused by the magnetic contribution to the free energy and, in essence, results from the barrierless character of the bcc-fcc transition if the system is forced to stay in ferromagnetic state, as was found earlier [22,28]. One can see in Fig. 1 that the differ-ence between the magnetic contribution to the energy and to the free energy (that is, the contribution of magnetic entropy) is essential. Neglect of the magnetic entropy would lead to essentially higher potential barriers.

We did not consider in Ref. [55] spatial scales related with the interphase fluctuations. Here we demonstrate that these fluctuations represent themselves the dynamic tweed structure with a typical inhomogeneous scale of the order of hundreds of interatomic distances. For the case of elemental iron, this is our prediction which has to be checked experimentally. However, for the alloy Fe-Ni which is frequently used as a model system structural transformations in iron and steel (due to the closeness of the transition temperature to room temperature), the developed short-range order at these spatial scales has been, probably, already observed [23,56]. To our knowledge, all previous discussions of these phenomena [13–15] where based on the concept of phonon soft mode which is not directly applied to fcc Fe (see the Introduction). Our model emphasizes the decisive role of the *magnetic* contribution to the free energy and its dependence on elastic deformations.

Energy relaxation of internal strains accompanying the transformation is the main driving force for the observed morphology of martensitic plates. In reality, some mechanisms of plastic accommodation of the internal deformations (creation and motion of dislocations) always take place, therefore, the observed structure turns out to be less regular. In our simulations we obtained an irregular pattern instead of twin plates at decrease of the parameters $A_v$, $A_s$, determining the elastic energy in Eq. (3). It is worth to notice that a similar pattern evolution takes place for magnetic domains in thin films and multilayers [45,47]. If the field of crystalline anisotropy directed perpendicular to the film plane is much larger than the in-plane component of external magnetic field regular magnetic stripes are observed. With the increase of the in-plane field the order parameter (normal component of magnetization) decreases and more isotropic and chaotic patterns reminiscent the case of suppressed elastic strains in our simulations.

To be able to compare our model with properties of real iron *quantitatively* the former should be modified, namely, phonon contributions to the free energy should be included. However, for relatively narrow temperature interval near the Curie temperature the phonon contributions are very weakly temperature dependent, at least, there are no serious theoretical or experimental reasons to assume the opposite. Contrary, in chromium- [57] or manganese- [58] based alloys phonon spectra are strongly temperature dependent near the Neel temperature. We assume this is not the case for the elemental iron.

Another important approximation is the two-dimensional character of the model proposed. A transition from two-dimensional to three-dimensional models was discussed for similar problems in Ref. [5]. The picture becomes more reach and complicated, however, qualitative conclusions such as appearance of inhomogeneity at the scale of tens of nanometers and associated slow fluctuation dynamics seem to be robust and reliable.

To conclude, we have demonstrated a crucial role of magnetoelastic phenomena, namely, a strong sensitivity of the potential relief along the Bain deformation path to magnetic state, for kinetics of the polymorphous transformation in Fe. Our model reproduces correctly the martensitic (shear) transition at low enough temperature and, at the same time, suggests a scenario of the massive transition at higher temperatures. The latter is developed by the nucleation and growth mechanism and includes spatial and temporal fluctuations of the order parameter (tetragonal deformation) with a characteristic tweed structure. It would be important to check this prediction by diffraction experiments at high temperatures.

## References.